# Instant single-pixel imaging: on-chip real-time implementation based on instant ghost imaging algorithm

Zhe Yang[1], Jun Liu[2], Wei-Xing Zhang[1], Dong Ruan[1], and Jun-Lin Li[1, *]

*1 State Key Laboratory of Low-dimensional Quantum Physics and Department of Physics, Tsinghua University, Beijing 100084, China*
*2 Wuhan Digital Engineering Institute, Wuhan 430074, China*
*\*center@mail.tsinghua.edu.cn*

**Abstract:** Single-pixel imaging (SPI) uses a single-pixel detector to create an image of an object. SPI relies on a computer to construct an image, thus increasing both the size and cost of SPI and limiting its application. We developed *instant single-pixel imaging* (ISPI), an on-chip SPI system that implements real-time imaging at a rate of 25 fps. ISPI uses the instant ghost imaging algorithm we proposed which leverages signal differences for image creation. It does not require a computer, which greatly reduces its cost and size. The reconstruct time of ISPI for image creation is almost zero because little processing is required after signal detection. ISPI paves the way for the practical application of SPI.

## 1. Introduction

The single-pixel imaging (SPI) process is to illuminate an object with structured light, collect transmitted or reflected light from the object using a single-pixel detector, and then perform calculations to create an image of the object [1,2]. Recently, SPI based on ghost imaging (GI) [3–11] and computational ghost imaging (CGI) [12–15] has attracted widespread attention. SPI is of particular interest in areas where array cameras are unavailable or expensive, such as terahertz imaging [16,17], infrared imaging [18] and X-ray imaging [19]. It also has important applications in the fields of three-dimensional imaging [20–23], lidar [24,25], microscopic imaging [26,27], and encryption [28,29]. Recently, encoding methods based on complete orthogonal bases [30,31], such as the Fourier basis [32–35] and the Hadamard basis [36–40], have significantly improved SPI image quality and imaging speed [41]. SPI can also produce high-quality images using the compressive sensing algorithm with sub-Nyquist sampling [42–45].

However, SPI requires a computer to create the image of the object. Image creation is time-consuming and requires considerable computing resources. This requirement constrains the use of single-pixel cameras in practice. Conventional digital cameras use chips for imaging (pictures taken on mobile phones provide a ubiquitous example), which suggests that reducing the dependence of SPI on computers is an important problem to address. There has been research implementing a field-programmable gate array (FPGA) on the SPI, but a computer was also required [46]. The recently developed instant ghost imaging (IGI) algorithm provides an on-chip GI [47–51], thereby making the computer redundant. We first proposed the IGI algorithm in 2015 [49]. Jun-Lin Li introduced the algorithm into computer-based experiments as sequence differential ghost imaging (SDGI) in his doctoral dissertation in 2016 [50]. The IGI algorithm is not application-specific and can be used in SPI to make it computer-independent.

In this paper, we developed instant single-pixel imaging (ISPI), an on-chip SPI system that implements real-time imaging for moving objects at a rate of 25 fps. The IGI algorithm uses signal differences between two consecutive temporal measurements for image reconstruction. The image reconstruction time of the IGI algorithm is almost zero, i.e., no post-processing is necessary. Therefore, ISPI does not require a computer because it has few computational resources demands, which greatly reduces both the size and cost of SPI.

Meanwhile, using signal differences is inherently noise-free (the noise components of two consecutive signals are approximately equal, therefore, taking the signal difference effectively removes the noise), which gives ISPI an excellent capacity for resisting interference due to optical background noise. This on-chip implementation of ISPI makes SPI more accessible for practical applications.

## 2. Method

### 2.1 Instant ghost imaging algorithm

The IGI algorithm takes the form [47,48]:

$$G^{IGI}(x) = \frac{1}{2N}\sum_{n=1}^{N}(S_{n+1} - S_n)[I_{n+1}(x) - I_n(x)] \quad (1)$$

where $S_{n+1} - S_n$ is the signal difference between two successive measurements detected by a single-pixel detector; $I_{n+1}(x) - I_n(x)$ is the signal difference between two successive patterns in SPI; and $N$ is the total number of measurements required to form an image.

We define the accumulation term $Acc_n^{IGI} = (1/2N)\sum_{i=1}^{n}(S_{i+1} - S_i)[I_{i+1}(x) - I_i(x)]$, where $n$ is the $n^{th}$ measurement and the index $i$ ranges from 1 to n. The IGI algorithm simultaneously measures and processes. When the last measurement has been made, the accumulation term is the resulting final image. The image calculation processing time is almost zero, so it is considered to be *instant*.

Furthermore, we gave a detailed proof in [47] that Eq. (1) is approximately equal to the standard background-subtraction-related imaging equation

$$G(x) = \langle (S - \langle S \rangle)[I(x) - \langle I(x) \rangle] \rangle \quad (2)$$

where the ensemble average is $\langle \cdot \rangle = (1/N)\sum_{i=1}^{N}(\cdot)$.

### 2.2 Experimental configuration

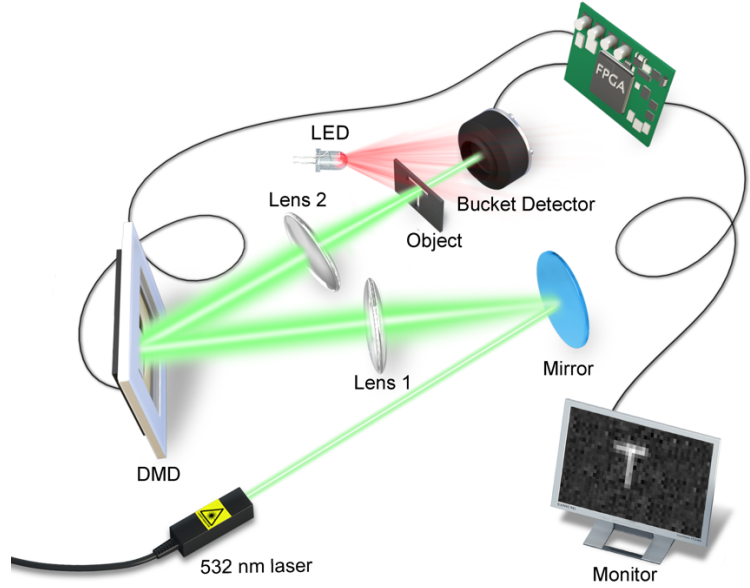

**Fig. 1.** The experimental setup of ISPI.

The equipment used in the experiment is shown in Fig. 1. A 532 nm reflected laser beam is expanded by lens 1. It illuminates the DMD, which operates at a frequency of 20 kHz, with random matrix coding to generate structured light. Lens 2 images DMD on the object. The square object contains a cut-out letter T, whose size is 0.8 mm × 0.8 mm. Transmitted light that passes through the T is collected by the bucket detector.

In the following, we analyze the workflow of ISPI. The FPGA (Xilinx XC7K325T) receives signal $S_n$ and calculates $S_{n+1} - S_n$ using signal $S_{n-1}$ stored in the inbuilt memory of the FPGA. When the calculation has been completed, $S_n$ overwrites $S_{n-1}$. During the same iteration, $I_n(x)$ and $I_{n-1}(x)$ are retrieved from FPGA memory to calculate $(S_n - S_{n-1})[I_n(x) - I_{n-1}(x)]$, which is added to $Acc_{n-1}^{IGI}$ (also stored in FPGA memory) to give $Acc_n^{IGI}$, and then $Acc_n^{IGI}$ overwrite $Acc_{n-1}^{IGI}$. When the last measurement is completed, the FPGA calculates $Acc_N^{IGI}$, which is the image of the object. Therefore, the image can be obtained in real time after the measurement, and the image reconstruction time is almost zero.

To demonstrate the ability of ISPI to resist interference from optical background noise, we placed an LED capable of producing light at different frequencies and intensities in the configuration to introduce light into the optical path as optical background noise.

### 3. Results

#### *3.1 Images of the object*

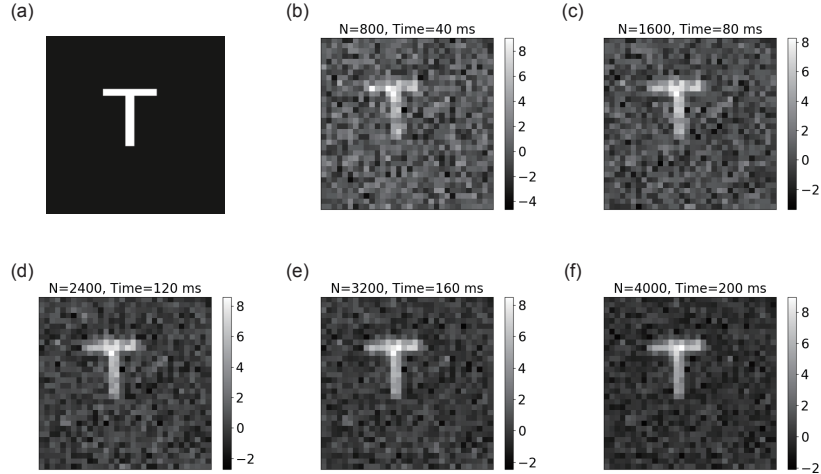

**Fig. 2.** Experimental results of ISPI.

Fig. 2(a) shows the object, the letter T. The size of the image is 32 pixels × 32 pixels. Figs. 2(b)–2(e) show the images produced by ISPI. The total number of measurements $N$ were respectively 800, 1600, 2400, 3200, and 4000. The results show that ISPI can create an accurate image of the object.

Note that the time on each image (the imaging time) is the time required for ISPI to generate an image when the DMD operated at 20 kHz. Because the ISPI image construction time is almost zero, and no post-processing is required, the imaging time is the time required for the DMD to project $N$ patterns.

### *3.2 Images of a moving object*

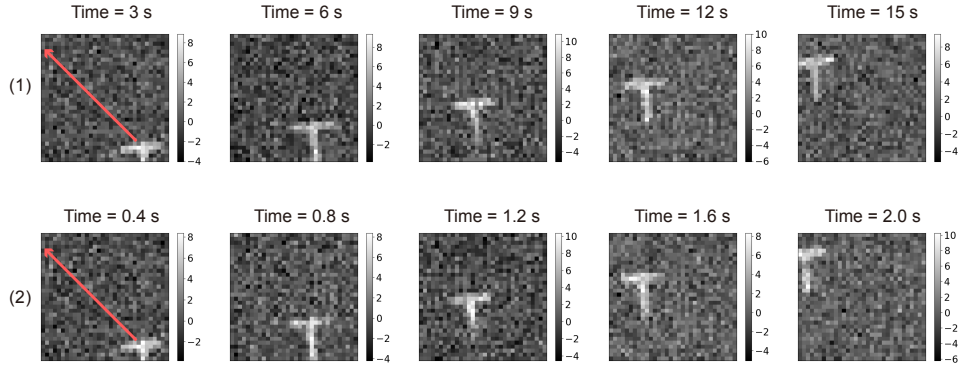

**Fig. 3.** Images of the moving object.

ISPI can create images of moving objects and produce real-time video at a frame rate of 25 fps. The frame rate is equal to $V_{DMD}/N$, where $V_{DMD} = 20$ kHz, which is the DMD operating frequency, and $N$ is the number of measurements required to create one image. In Figs. 3, we set $N = 800$, and the frame rate was 25 Hz, which is better than the standard video rate of 24 fps. As shown in the leftmost figure of each row, the object moved from the lower right to the upper left. The speed at which the object moved was controlled by a stepper motor. In row (1) of Fig. 3, the object had a speed of 0.1 mm/s; in row (2), it had a speed of 0.8 mm/s. Since ISPI does not need to perform post-processing calculations, it can produce real-time videos. The related videos of Figs. 3 are shown in Visualizations 1 and 2.

### 3.3 Resistance to optical background noise

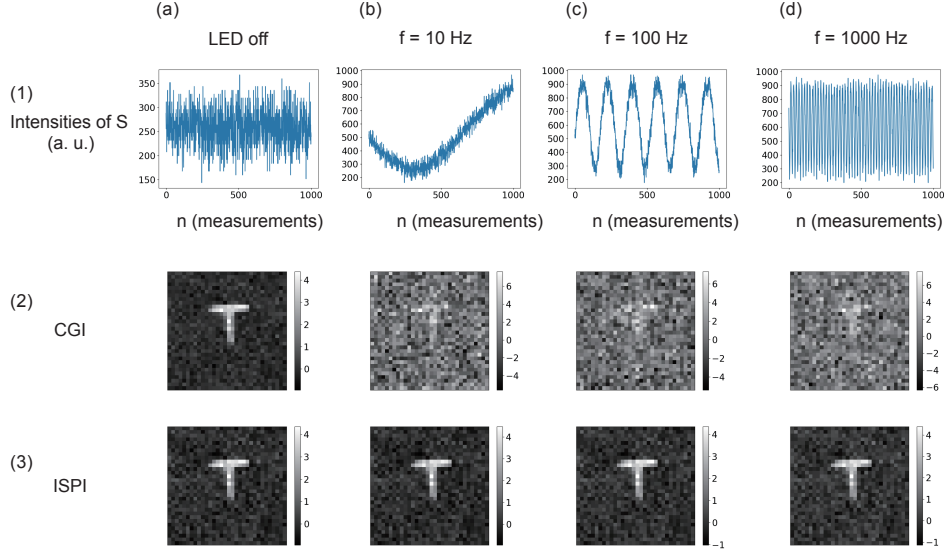

**Fig. 4.** The resistance of ISPI to optical background noise.

The signal difference algorithm inherently resists common noise modes. To show that ISPI resists optical background noise, we conducted an experiment in which an LED was used to introduce background noise to the laser signal. The LED was controlled to generate optical background noise of different intensities and different frequencies (Fig. 4).

In Fig. 4, row (1) is the signal curve of the bucket detector $S$ for different noise frequencies. Row (2) shows the images produced by a conventional CGI algorithm as Eq. (2). Row (3) shows the results given by ISPI. The total number of imaging measurements was $N = 4000$. Column (a) shows the bucket detector signal $S$ and images obtained with no added background noise, and columns (b)–(d) show the bucket detector signal $S$ and images obtained with different optical background noise operating frequencies.

It can be seen from row (1) that the background noise has a great effect on the measurements made at the barrel detector. Row (2) shows that image quality significantly decreased when the CGI algorithm was subjected to background noise, even to the extent of no image being produced. Row (3) shows that ISPI is extremely resistant to optical background noise.

Furthermore, the ISPI can well resist the optical background noise when

$$(Q_{n+1} - Q_n) \ll (S_{n+1} - S_n) \tag{3}$$

where $Q$ is the intensity of optical background noise. The optical background noise is required to be a slow variation signal compared to $S$. If this condition is satisfied, the ISPI will not be affected by the optical background noise regardless of the type such as Poisson noise, Gaussian noise.

### 4. Discussion and Conclusion

In this paper, we described ISPI, an on-chip SPI using the IGI algorithm. Although the previous work has involved the utilization of an FPGA in the field of GI [47], we are the first to implement high-speed real-time SPI that is truly computer-independent. We are able to accomplish this because the IGI algorithm requires few computational resources for image creation. This implementation greatly reduces both the size and cost of SPI, which makes SPI more practical as it aligns the SPI with current digital cameras and phones that require only a chip for imaging and have no need for a computer.

The image creation time of ISPI is almost zero, and the calculation required is completed at the time the image is acquired. Therefore, there is no need for any post-processing. Thus, ISPI is effectively a real-time imaging system, which is why we named it *instant*.

Another advantage of ISPI is its resistance to optical background noise. SPI and CGI can both resist turbulence and scattering medium [52–54] but do not have a strong capacity for resisting optical background noise [55,56]. We have experimentally verified that ISPI has excellent resistance to optical background noise. In an environment containing optical background noise that changed drastically in intensity, ISPI was almost unaffected by the noise and produced an accurate image of the object because the signal difference algorithm intrinsically resists common noise modes [48].

The ISPI can be improved in three areas: imaging speed, imaging quality, and image size.

### Imaging speed

ISPI currently operates at 20 kHz. This operating frequency is governed by the DMD, not by any part of the ISPI firmware. The FPGA can operate at >100 MHz, making it potentially 5000 times faster than it is at the current operating frequency. Thus, the imaging speed of ISPI can be greatly increased by using faster DMDs or faster light source arrays [41].

### Imaging quality

In this work, a random basis with 0 and 1 was used for imaging. It is well known that a random basis cannot be used for high-definition imaging, which uses artifacts such as the orthogonal basis [30–40]. ISPI using a complete orthogonal basis and the signal difference algorithm are being studied.

### Image size

The ISPI images are currently limited by the internal memory space of the FPGA chip. We plan to augment the FPGA memory with external double data rate-random access memory (DDR-RAM) to create larger images of 1024 pixels × 768 pixels.

In summary, we prototyped ISPI, an on-chip high-speed real-time SPI using the IGI algorithm. ISPI can image objects without post-processing, can produce real-time images of moving objects, and has excellent resistance to optical background noise. It also greatly reduces the cost and size of SPI, reducing reconstruction time to almost zero. We believe this application will greatly promote the practicality of SPI.

### Funding

National Natural Science Foundation of China (NSFC) (51727805).

### Disclosures

The authors declare no conflicts of interest.